# Fundamental miniaturization limits for MOSFETs with a monolayer MoS$_2$ channel

**Maksym V. Strikha[1, 2], Mykola Yelisieiev[3,*], and Anna N. Morozovska[4,†],**

[1] Taras Shevchenko Kyiv National University, Faculty of Radiophysics, Electronics and Computer Systems, pr. Akademika Hlushkova 4g, 03022 Kyiv, Ukraine

[2] V. Lashkariov Institute of Semiconductor Physics, National Academy of Sciences of Ukraine, pr. Nauky 41, 03028 Kyiv, Ukraine

[3] Taras Shevchenko National University of Kyiv, Faculty of Physics, pr. Akademika Hlushkova 4a, 03022 Kyiv, Ukraine

[4] Institute of Physics, National Academy of Sciences of Ukraine, pr. Nauky 46, 03028 Kyiv, Ukraine

**Abstract**

We propose a theoretical model for describing the operation of a transistor with a MoS$_2$ monolayer channel, which allows to obtain an analytical approximation of the potential in the channel. This potential depends on the drain and gate voltages. On this basis we make estimates for the minimum channel lengths due to the fundamental restriction of quantum tunneling through the barrier. It is shown that the relatively large effective mass of electrons in the MoS$_2$ monolayer allows to predict the creation of devices with channels of a significantly shorter (2.5 - 3 nm) length than in traditional silicon MOSFETs. These devices can be promising for the ultra-fast electronics of new generation.

## I. Introduction

Hopes for the emergence of mass carbon electronics with unique high-frequency characteristics, which appeared immediately after the experimental production of graphene [1], did not come true due to the gapless nature of the band structure of graphene, which made it difficult to obtain two distinct states "0" and "1" (see Ref. [2]). The creation of an energy gap of the order of 100 meV in the band spectrum of graphene in that or this way (transition to the nanoribbons, surface hydrogenation, etc.) immediately reduced the mobility of electrons to values of the order of silicon and below [3, 4]. This led to intensive studies of MOSFET transistors based on 2D (single- or multilayer) transition metal dichalcogenides (TMD), which are direct-band semiconductors with a sufficiently wide band gap [5].

---

[*] a.k.a. − Nicholas E. Eliseev
[†] Corresponding author: anna.n.morozovska@gmail.com

The 2H MoS$_2$ monolayer has a 1.8 eV bandgap, though it is characterized by low carrier mobility [of order of см$^2$/(V s)] on a standard SiO$_2$ substrate, which makes it unpromising for electronic implications. However, the use of a HfO$_2$ substrate with high values of the dielectric permittivity (approximately equal to 25) allows to suppress the scattering of carriers on the ionized impurities of the substrate and increase the value of the mobility to ~ 200 см$^2$/(V s). Due to this an effective transistor based on a MoS$_2$ monolayer was created [6]. It has a channel length of $L = 1.2$ μm and the current relation in the ON\OFF modes approximately equal to $10^8$.

Later on [7] a transistor on a MoS$_2$ monolayer with a channel length of 7.5 nm was created, although the level of its functionality was low due to the strong effect of the drain-induced barrier lowering (DIBL) and direct tunneling of carriers through the barrier; however, the transistor with $L = 90$ nm studied in the same work had operated with a sufficient level of functionality. In recent years, 2D MoS$_2$ transistors have also been used to create memory cells. Initially, an organic relaxor was used as a ferroelectric substrate [8]. However, this required high switching voltages (~ 20-30 V) to switch between the two states of a memory cell. The transition to Hf$_{1-x}$Zr$_x$O$_2$ ferroelectric [9] has allowed to reduce this value by an order of magnitude and to discuss the practical use of such memory cells in modern electronics devices. A bit later, more MOSFETs with ultra-short (sub-10 nm [10] and 3.8 nm [11]) lengths of MoS$_2$ channels have been realized on different substrates. Mention, that a SiO$_2$ substrate, used in [11], made the device, despite of its short channel, unpromising for applications in electronics because of its low electron mobility (~20 см$^2$/(V s)). Therefore HfO$_2$ substrate can be treated as a most perspective one due to reasons discussed above.

However, a consistent theoretical description of such systems is still virtually absent. To model them, previous authors used either a standard virtual source (VS) model developed for silicon field-effect transistors (see, for example, [12]), or its improved "Massachusetts" version (MVS) [13], which provided for the adjustment of theoretical curves to experimental data by appropriate selection of a number of parameters.

In this paper, we try to construct a theoretical approach for answering the following question. What is the minimum channel length that we can expect for transistors based on the MoS$_2$ monolayer? This issue is especially relevant, because after reaching the working lengths of the channel of 5 nm order, the possibility of a traditional silicon MOSFET miniaturization is almost exhausted [14, 15].

## II. Theoretical approach

Below we consider the circuit of a field-effect transistor on a MoS$_2$ monolayer implemented in Ref.[7]: the conductive channel lies on a layer of HfO$_2$ oxide, which in turn is placed on a silicon substrate, which acts as a back gate. The source (S) and drain (D) contacts are metal conductors. In real devices, different metals are used for this: Au [6] or Ni [9]. Almost ideal Ohmic contacts were implied in [7] on the connection of metallic and semiconductor modifications of a monolayer MoS₂, which are characterized by almost identical work function/electronic affinity, so the Schottky barrier (even if any arises at the contact) can be neglected. The band structure of such a device is shown in **Fig. 1a**. Let us underline that it differs significantly from the band scheme of the traditional silicon MOSFET shown in **Fig. 1b**.

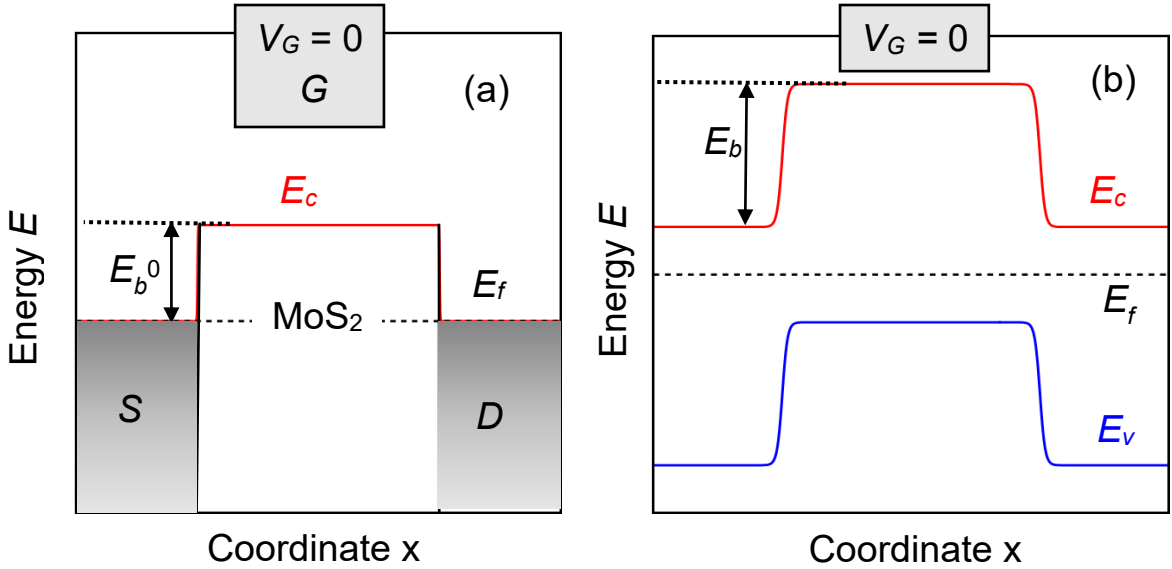

**FIGURE 1.** MOSFET on a monolayer MoS₂ **(a)** and a traditional silicon MOSFET **(b)** for the case of zero gate voltage, $V_G = 0$.

According to **Fig. 1b** and the standard theory of semiconductors is (see, e.g, [16]), the height of the potential barrier between the source and drain at zero voltage at all terminals of the silicon MOSFET transistor can be written as follows:

$$E_b = k_B T \ln\left(\frac{N_A N_D}{n_i^2}\right). \tag{1}$$

Here $N_D$ and $N_A$ are the concentrations of donors and acceptors, $n_i$ is the concentration of intrinsic carriers for the temperature $T$, and $k_B$ is a Boltzman constant. We consider the non-degenerated carriers, and regard that donors and acceptors are completely ionized. The order of $E_b$ magnitude is equal to the band gap.

On the other hand, for the case of MOSFET on the MoS₂ monolayer (**Fig. 1a**), the height of the barrier is determined by the Fermi energy level relative to the bottom of the conduction band. In the same approximation of nondegenerate carriers and fully ionized donors, it is equal to:

$$E_b^0 = k_B T \ln\left(\frac{N_c}{N_D}\right). \tag{2}$$

Here $N_c$ is the effective states density in the conduction band. For the 2D MoS₂ it is given by the standard expression:

$$N_c = \frac{k_B T m_c}{\pi \hbar^2}. \qquad (3)$$

The expression (2) also includes the 2D concentration of ionized donors in the MoS$_2$ channel. This value can be estimated based on the data from the work [7].

The situation shown in **Fig. 2a** corresponds to the "ON" state of the transistor, when free electrons are in the channel. In particular, the accumulation mode, when the number of free electrons exceeds the number of ionized donors, is shown. In order to switch the transistor to "OFF" state, it is necessary to apply a negative threshold voltage $V_T$ to the gate "G", sufficient to switch the channel to depletion mode, when the free electrons in the channel are no longer present and the space charge density is determined by ionized donors only (**Fig. 2b**).

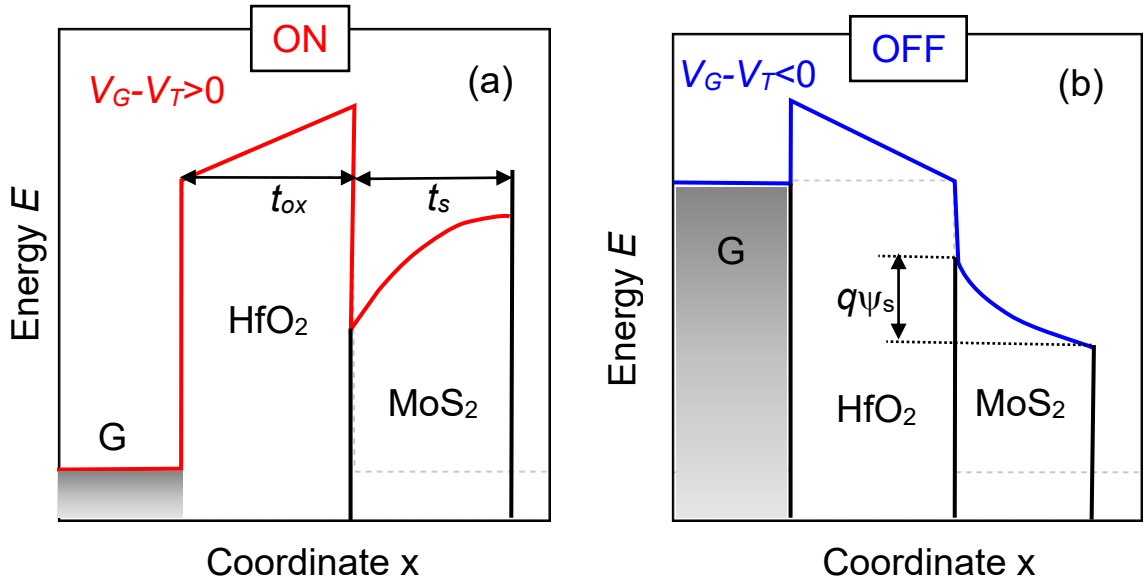

**FIGURE 2.** "ON" (a) and "OFF" (b) states of MOSFET on MoS$_2$ monolayer.

In the general case, the value of current $I_{SD}$ in the source-drain circuit is determined by the product of the channel width, $W$, 2D charge density of the free electrons at the selected point of the channel (for which often the so-called "virtual source", i.e., the point of the top of the potential barrier is chosen), $Q(0)$, and the average electron velocity at the same point $\langle v(0) \rangle$:

$$I_{SD} = W Q(0) \langle v(0) \rangle. \qquad (4)$$

Considering the conducting channel of the transistor and its gate as the plates of a flat capacitor, we assume that $Q(0)$ can be estimated for the gate voltage higher than the threshold voltage as:

$$Q(0) = C_{ox}(V_G - V_T). \qquad (5)$$

The capacitance $C_{ox}$ is equal to

$$C_{ox} = \frac{\varepsilon_{ox} \varepsilon_0}{t_{ox}}. \qquad (6)$$

The capacitance $C_{ox}$ depends on the oxide layer thickness $t_{ox}$ and its dielectric permittivity $\varepsilon_{ox}$ ($\varepsilon_0$ is a universal dielectric constant).

Considering that $Q(0) = qN_D$ in the depletion mode, and using the values $\varepsilon_{ox} = 25, t_{ox} = 26$ nm, $V_T = 0{,}5$V from Ref. [7] (see also **Fig. 3a** therein), we obtain an estimation for the concentration of donors, $N_D = 2.5 \times 10^{16} \text{m}^{-2}$, which coincides in the order of magnitude with the values obtained by other methods for a MoS$_2$ monolayer.

When modeling real current-voltage (I-V) curves in nanotransistors, there is a significant problem in determining the value $\langle v(0) \rangle$ depending on the source-drain voltage $V_{DS}$. It is generally accepted for traditional silicon-based MOSFETs that

$$\langle v(0) \rangle = \frac{\mu V_{DS}}{L} \tag{7}$$

for the linear mode in long-channel transistors at low values of drain voltages. Here $\mu$ is the electron mobility. For high field values the electron velocity is assumed to be saturated due to intensive scattering,

$$\langle v(0) \rangle = v_{sat}. \tag{8}$$

The same approach is usually applied to the transistors with a MoS$_2$ monolayer channel. Thus, it was obtained in Ref.[17] that $v_{sat} = (3.4 \pm 0.4) \times 10^6$ m/s for electric fields higher than 3-4 V/μm in a transistor with $L = 4.9$ μm. However, as it will be shown below, such an approximation acceptable for a traditional silicon MOSFETs based on a SiO$_2$ oxide layer with a low dielectric constant ($\varepsilon_s = 3.9$), should be used with precaution for MOSFETs with MoS$_2$ monolayer channel with $\varepsilon_s = (4.2 - 7.6)$ [18], and HfO$_2$ oxide layer with almost an order of magnitude higher dielectric constant $\varepsilon_s = 25$. In fact, as is shown by the solution of the two-dimensional Poisson equation [16], the channel depth for which the drain potential perturbs the potential in the channel is described by the parameter

$$\Lambda = \sqrt{\frac{\varepsilon_s}{\varepsilon_{ox}} t_s t_{ox}}, \tag{9}$$

which depends on the semiconductor layer thickness $t_s$ and its dielectric permittivity $\varepsilon_s$. Later we take $\varepsilon_s = 5$ and $t_s = 0.7$ nm for the considered MoS$_2$ monolayer [19]. It is seen from Eq.(9) that the perturbation penetrates into the channel only at a distance $\Lambda \sim (1 - 2)$ nm for the gate oxide parameters $\varepsilon_{ox} = 25, t_{ox} = (10 - 30)$ nm, respectively. Therefore, an extremely effective screening occurs in the channel and reduces the undesirable DIBL effect and therefore decreases the subthreshold swing (SS) [16, 20]. The $\Lambda$ value for a silicon channel with a SiO$_2$ oxide layer is significantly higher.

Thus, the analytical form of potential in the channel can be approximated by a step-linear function:

$$U(x) = \begin{cases} E_b^0(V_G), & 0 < x \leq L - \Lambda \\ E_b^0(V_G) - [E_b^0(V_G) + qV_{DS}]\frac{x-L+\Lambda}{\Lambda}, & L - \Lambda < x \leq L \end{cases} \tag{10}$$

Hereinafter we regard that $qV_{DS} \geq 0$, where $q$ is an electron charge, and, naturally the barrier $E_b^0 \geq 0$. The function (10) is shown in **Fig. 3**. $E(x)$

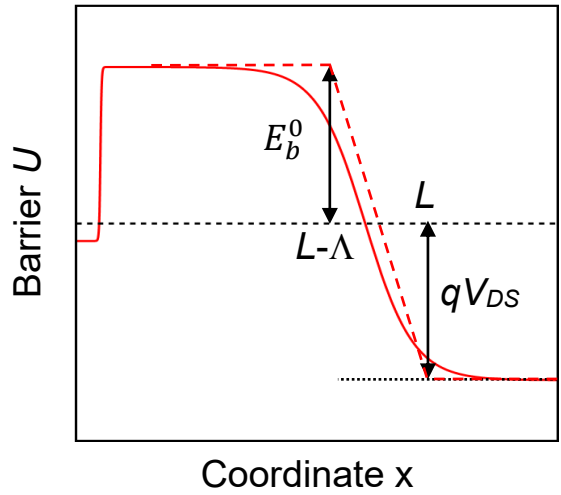

**FIGURE 3**. The model potential $V(x)$ in the MoS$_2$ monolayer channel. The voltage $V_{DS}$ is applied between the source and the drain.

This shows that even for long-channel transistors with a MoS$_2$ monolayer channel and an HfO$_2$ oxide layer, the electron transport through this channel must be significantly different from the transport in the channel of silicon MOSFET. In the latter case (this is confirmed, in particular, by the results of numerical simulations [21]) in the region of the virtual source near the top of the barrier there is a relatively short "nanochannel" with low electric fields. Instead, most of the channel toward the drain is characterized by sufficiently high fields, which determine the applicability of Eq.(7). In contrast, for transistors with a relatively long MoS$_2$ monolayer channel and a HfO$_2$ oxide layer, the field is small along almost the entire channel length, so that the acceleration of electrons by a strong field occurs only in a relatively narrow region of the "cut-off" near the drain, with the thickness described by a screening length Λ. Therefore, even the use of standard expressions (7)-(8) for such transistors requires separate verification. For the reasons, described above, it is clear, that MVS model, elaborated for silicon MOSFETs [13, 16], can fail in description of transistors with a MoS$_2$ monolayer channel. At the same time, this kind of highly screened potential in the channel makes transistors with a MoS$_2$ monolayer channel and a HfO$_2$ oxide layer very promising for obtaining remarkably short channels.

In fact, the results obtained in [14] are consistent with today's generally accepted notion that the classical model of MOSFET transport is quite applicable to Si transistors with a conduction channel length of up to 10 nm and even slightly less. However, further scaling to 5 nm already poses serious problems, both applied (increasing the role of parasitic resistances and capacitances with very short conduction channels) and fundamental - due to the tunneling of electrons through the barrier. Taking into account the realistic potential in the channel shows, that the electron tunnels through a region significantly shorter than the physical length of the channel $L$ under the drain voltage, and therefore an available estimate of the minimum quantum constraint silicon MOSFET $L_{min} \approx 1.2$ nm (see e.g., [16]) is significantly underestimated. From here it is clear why after reaching the 5 nm working channel lengths it was not possible to reach the declared 3 nm values [22] maintaining the proper level of functionality of the transistor.

Instead, the form of the potential (10) shows that the drain voltage variation slightly changes the width of the tunneling region for the electron for the case $L \gg \Lambda$. However, this conclusion, as will be shown below, is no longer applicable to ultrashort channels with $L \sim \Lambda$. In this case, the tunneling area can be very significantly reduced with increasing of the drain voltage. In general, the value of the effective electron mass for the MoS$_2$ monolayer [19] equals to $m_c = 0.55 m_o$, where $m_o$ is the free electron mass. This can lead to a much shorter channel length than in a traditional MOSFET.

As it is well known, there is a fundamental limit for a minimal barrier height in the channel at which the transistor retains its functionality and can operate in physically distinct ON\OFF states [23]:

$$E_{min} = k_B T \ln 2. \qquad (11)$$

For room temperature it equals to 17 meV. For the above value of the concentration of ionized donors, the barrier height $E_b^0$ in the transistor with the MoS$_2$ monolayer channel described by expression (2) is 25 meV, so from this point of view, the considered device is a functional one.

Within the framework of the same simple scheme, it is possible to estimate the minimum length of the MOSFET conduction channel. The height of the barrier in the "OFF" mode must be at least not less than $E_{min}$, which guarantees that the electrons, when they overcome the barrier, do this with a probability $P$ less than 1/2. In this case, the minimum width of the barrier (channel length) is determined by quantum mechanical tunneling through the barrier. The probability that an electron from a source tunnels through a barrier can be estimated in the Wentzel-Kramer-Brillouin (WKB) approximation, which gives a well-known formula for the tunneling probability of a particle with energy $E$ and mass $m_c$ through a barrier with potential $U(x)$ between the turning points $x_1$ and $x_2$:

$$P \approx \exp\left(-\frac{2}{\hbar}\int_{x_1}^{x_2} \sqrt{2m_c(U(x) - E)} dx\right) \qquad (12)$$

We first consider the potential of the barrier to be rectangular along the entire channel length $L$. Let us set its height $U(x)$ equal to the minimum value $E_{min}$ described by Eq.(11) for $x_1 = 0$ and $x_2 = L$. Hence for the minimum length of the conduction channel, corresponding to $P = 0.5$, we obtain

$$L_{min} = \frac{\ln 2}{2} \frac{\hbar}{\sqrt{2m_c(E_{min} - E)}}. \qquad (13)$$

Putting $E = 0$, we obtain that $L_{min} = 0.7$ nm. However, for such $L_{min} \leq \Lambda$ the source is no longer electrically screened from the drain, and therefore we can use a linear formula (as the first approximation) along the entire channel:

$$U(x) = E_{min} - \frac{x}{L}(E_{min} + qV_{DS}) \qquad (14)$$

As it follows from the expression (14), the application of a minimum positive voltage $\frac{E_{min}}{q} \cong 17$ mV to the drain reduces the tunneling length for an electron injected with Fermi energy of the source for a device operating at room temperature, and leads to a corresponding increase in the minimum channel length.

Substituting Eq.(14) to Eq.(12) at $E = 0$, we obtain the value of the minimum channel length, which depends on the drain voltage:

$$L_{min} = \frac{\ln 2}{2} \frac{\hbar \chi}{\sqrt{2 m_c E_{min}}}, \qquad \chi = \frac{3}{2}\left(1 + \frac{qV_{DS}}{E_{min}}\right) \qquad (15)$$

Note that Eq.(15) does not give an exact limiting case of Eq.(13) when the drain voltage tends to zero, because we assumed that the potential (14) no longer has a rectangular shape even at minimum drain voltages, but has a triangular shape. The use of exact self-consistent solutions of the Poisson equation for low voltages should eliminate this discrepancy.

Also note that the real probability of tunneling (12) is a function of the gate and drain voltages. In the general case, for the gate voltage, which significantly exceeds the threshold, and the positive drain voltage, we can use the expression

$$E_b(V_G, V_{DS}) = E_b^0 - \alpha(V_G - V_T) - \beta V_{DS}, \qquad (16)$$

where the first term in the right part is described by Eq.(2), $E_b^0 \cong k_B T \ln\left(\frac{N_c}{N_D}\right)$, the second one describes the change of the barrier height under applying the gate voltage, and the third term describes the decrease of the barrier height due to the DIBL effect. The coefficient $\alpha \ll 1$ due to the fact that the most part of the gate voltage drops in the oxide layer, while the coefficient also depends on $V_G$ in a general case. The coefficient $\beta$ is close to zero for a channel with a sufficiently small screening length (9). Note that Eq.(16) has a physical sense until $E_b \geq 0$, otherwise FET is in the state "ON" and gate voltage variation modulates the current in source-drain circuit no longer.

Let's try to estimate the coefficient α for the depletion mode. The Poisson equation for the semiconductor layer is written for this mode as follows:

$$\varepsilon_s \varepsilon_0 \frac{d^2 \psi(y)}{dy^2} = -\frac{qN_D}{t_s}. \qquad (17)$$

Here $y$ is coordinate, normal to the semiconductor-oxide interface. The solution of this equation with boundary conditions of zero potential and field at the semiconductor-vacuum boundary $y = t_s$ gives the following expression for the magnitude of the potential at the semiconductor-oxide interface (**Fig. 2b**):

$$\psi_s = \frac{qN_D t_s}{2\varepsilon_s \varepsilon_0}. \qquad (18)$$

The form of the solution (18) corresponds to the known solution for the Schottky barrier height, where, however, the Schottky layer width is replaced by the fixed thickness of the MoS$_2$ monolayer, which is orders of magnitude smaller. Substitution of the above parameters in Eq.(18) gives $\psi_s = 0.031$V at $\varepsilon_s = 5$ and $t_s = 0.7$ nm.

The voltage drop on the oxide layer equals to

$$\Delta V_{ox} = E_{ox} t_{ox}. \qquad (19)$$

Using the Gauss theorem for the field in the oxide layer, we finally obtain the following expression:

$$\Delta V_{ox} = \frac{qN_D t_{ox}}{\varepsilon_{ox}\varepsilon_o}. \qquad (20)$$

Substituting the above parameters to Eq.(20), we obtain that $\Delta V_{ox} = (0.2 - 0.46)$ V at $t_{ox} = (10 - 30)$ nm. Using the voltage of the flat bands $V_{FB} = 0.25$V [16], caused by a small difference in the electron affinity in silicon and molybdenum bisulfate ($\chi_{Si} = 4.05$eV and $\chi_{MoS_2} = 4.3$eV), and without taking into account the bound charge at the semiconductor-oxide interface, the estimative expression for $\alpha$ is $\alpha \approx \frac{\psi_S}{V_{FB} + \Delta V_{ox} + \psi_S}$. From the expression $\alpha \approx 0.04$ for $t_{ox} = 30$ nm and $0.07$ for $t_{ox} = 10$ nm.

Let us underline that the $\alpha$ values are significantly overestimated in comparison to the experiments [7], where the gate voltage controls the MOSFET current at $V_{DS} = 1$ V and gate voltage $V_{GT} = 2$V (hereinafter the voltage difference $V_{GT} = V_G - V_T$ is introduced). The reason is that it is impossible to obtain a simple analytical solution of Poisson equation for the case of the accumulation mode (**Fig. 2a**), but conventional considerations similar to those in Refs.[12, 16] allow us to conclude that α value is even smaller in this case, and in the accumulation mode the increasing the voltage on the gate leads only to a very slight additional bending of the zones down.

### III. Results and discussion

The voltage dependences of transparency coefficient $P$, shown in **Fig. 4**, are calculated for $t_{ox} = 10$ nm corresponding to $\Lambda = 1.18$ nm and $\alpha = 0.07$. The dependence of the transparency $P$ on the drain-source voltage $V_{DS}$ at fixed gate voltage $V_{GT} = -0.5\ V$ (dashed curves), 0 (solid curves) and 0.31 V (dotted curves) is shown in **Fig. 4a.** The dependence of the transparency $P$ on the gate voltage $V_{GT}$ at fixed $V_{DS} = 1\ V$ is shown in **Fig. 4b**. Different color of the curves corresponds to the channel length $L = 2, 2.5, 3, 3.5$ and 4 nm. It is seen from the curves, that the transparency monotonically increases with $V_{DS}$ increase for all $L$, and the increase prevents the controllable functionalization of the FET for short $L \leq 2$ nm. Longer channels ($L \geq 2.5$ nm) can be controlled by the gate voltage $V_{GT} \leq 0\ V$ at the drain voltages $V_{DS} \leq 1\ V$. Negative voltages $V_{GT}$ are required to close the MOSFET with a channel shorter that 3 nm at the drain voltage $V_{DS} = 1\ V$. A 4-nm channel is closed even at $V_{GT} = 0$, and a small positive gate voltage should be applied to open it. Anyway, the condition $L > 2\Lambda$ is required for the minimal gate control of MoS$_2$ channel at the level $P < 0.5$, and the condition $L > 3\Lambda$ provides the high gate control at the level $P < 0.1$. Note that the voltage dependences of transparency calculated for thicker oxide layers $t_{ox} = 20$ nm and 30 nm, shown in **Figs. A1b-c** and **A1e-f**, illustrate the same trends as in **Figs. 4a-b**, but higher $L \sim (3 - 5)$nm are required to provide the MOSFET operation in a gate-controllable ON/OFF states.

The color map of the transparency vs. the drain-source and gate voltages, $V_{DS}$ and $V_{GT}$, and $L = 3$ nm is shown in **Fig. 4c.** The color map of the transparency vs. the gate voltage $V_{GT}$ and channel length $L$ is shown in **Fig. 4c** for $V_{DS} = 1$ V**.** Dashed lines $P = 1/2$, which are related with quantum tunneling, separate the functional and non-functional modes of the FET; and solid lines $V_{GT} = 0$ separate its ON and OFF states with and without "classical" free carriers, respectively. Note that the color maps of transparency calculated for thicker oxide layers $t_{ox} = 20$ nm and 30 nm, shown in **Figs. A2b-c** and **A2e-f**, illustrate the same trends as in **Figs. 4c-d**, but higher $L \sim (3-5)$nm are required to provide the operation of MoS$_2$ channel in well-controllable ON and OFF states.

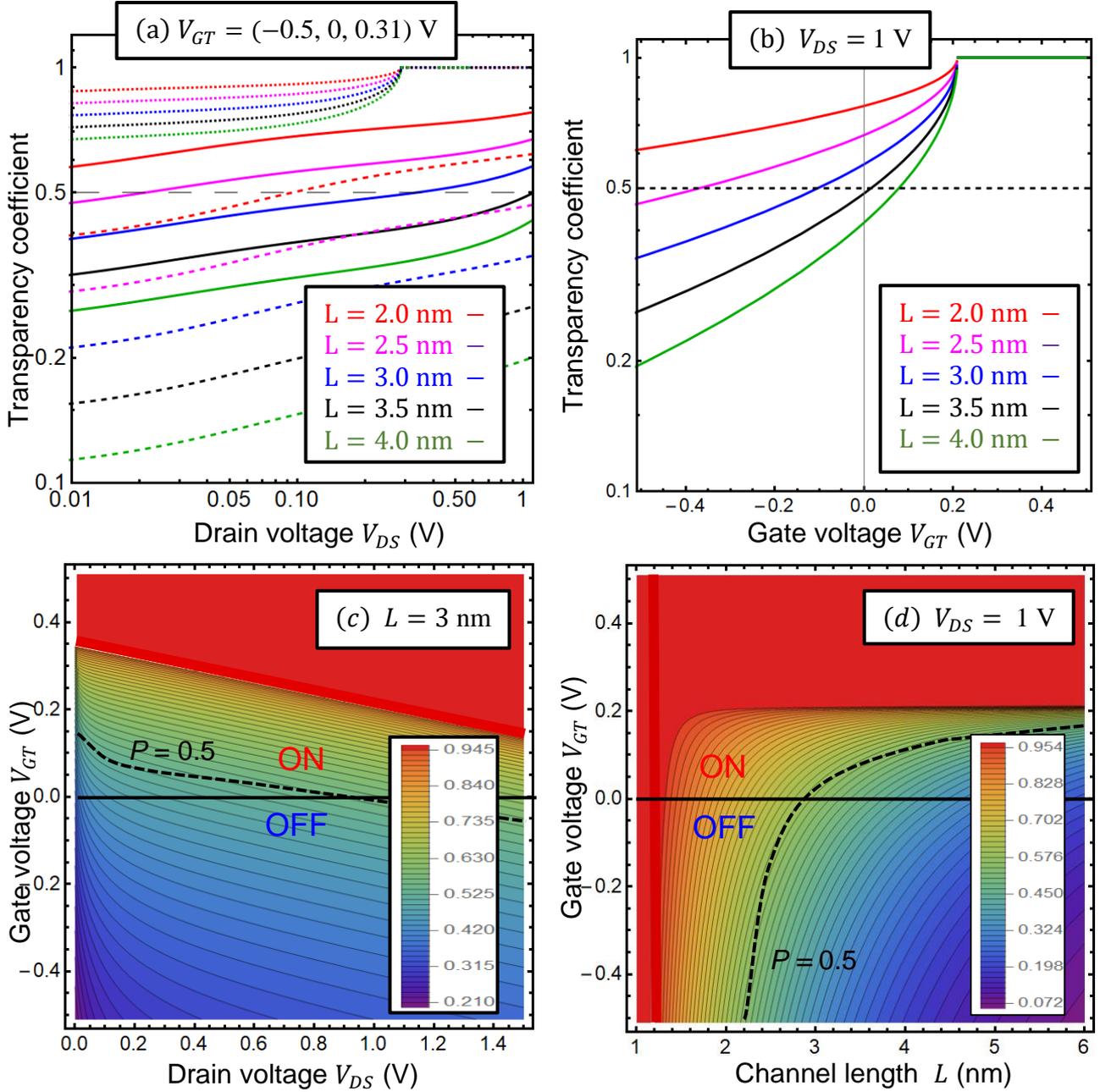

**Figure 4.** (a) The dependence of the transparency coefficient $P$ on the drain-source voltage $V_{DS}$ at fixed gate voltage $V_{GT} = -0.5$ V (dashed curves), 0 (solid curves) and 0.31 V (dotted curves). (b) The transparency $P$ on $V_{GT}$ at $V_{DS} = 1$ V. Different curves correspond to the channel length $L = (2-4)$nm (as listed in the legends). (c) The transparency $P$ vs. the voltages $V_{DS}$ and $V_{GT}$ at $L = 3$ nm. (d) The transparency $P$ vs. the gate voltage $V_{GT}$ and length $L$ at $V_{DS} = 1$ V. Dashed lines $P = 1/2$ separate the functional and non-functional modes of the FET; solid lines $V_{GT} = 0$ separate its ON and OFF states, respectively. Parameters $t_{ox} = 10$ nm, $\Lambda = 1.18$ nm, $\alpha = 0.07, \beta = 0.01, t_s = 0.7$ nm, $V_{FB} = 250$mV, $\psi_s = 30$mV, $\Delta V_{ox} = 200$ mV, $E_b^0 = 25$ meV, $\varepsilon_{ox} = 25$, $\varepsilon_s = 5$, and $T = 293$ K.

The main conclusion following from our analytical calculations and **Figs. 4**, **A1-A2** is that the monolayer MoS$_2$ channels as short as (2 – 3) nm in complex with gate oxide having high dielectric permittivity $\varepsilon_{ox} \geq 25$ (such as HfO$_2$ or other high-k ferroelectrics) can be used in optimized MOSFETs of the new generation. Obviously, there are many technical problems on the way of fabrication of such MOSFETs. However, the relatively large effective mass of electrons in the MoS$_2$ monolayer makes the fundamental restrictions more favorable allows us to hope for the creation of devices with channels with a significantly shorter (2 – 3 nm) length than in traditional silicon MOSFETs. Moreover, we can conclude that a functionality level of planar MOSFETs with a monolayer MoS$_2$ channel of the order of 1.5 nm corresponds to the functionality level of FET with CNT channel of the order of 7 nm according to the numerical modelling [24].

**Author's contribution.** M.V.S. and A.N.M. generated the research idea. M.V.S. performed most of analytical calculations and wrote the manuscript draft. M.Y. and A.N.M. performed numerical calculations, prepared graphics and improved the manuscript.

**Acknowledgements.** A.N.M. acknowledges the National Research Foundation of Ukraine (Project number 2020.02/0027).

### APPENDIX A. Transparency coefficient calculations

The transparency coefficient in VKB approximation has the form:

$$P \approx \exp\left(-\frac{2}{\hbar}\int_{x_1}^{x_2}\sqrt{2m_c(U(x)-E)}dx\right). \tag{A.1}$$

The potential barrier for a short channel is

$$V(x, V_{DS}) = E_{min} - \frac{x}{L}(E_{min} + qV_{DS}), \tag{A.2}$$

where $V_{DS}$ is a drain-source voltage, $E_{min} = kT \ln 2$, $q$ is an elementary charge, $L$ is the channel length. The potential barrier for a long channel ($L \gg \Lambda$) is

$$U(x) = \begin{cases} E_b(V_G), & 0 < x \leq L - \Lambda, \\ E_b(V_G) - \frac{E_b(V_G)+qV_{DS}}{\Lambda}(x - L + \Lambda), & L - \Lambda < x \leq L, \end{cases} \tag{A.3}$$

where $V_{DS}$ is a gate voltage and the screening length is $\Lambda = \sqrt{\frac{\varepsilon_s}{\varepsilon_{ox}}t_s t_{ox}}$. The barrier height is:

$$E_b = q(V_b^0 - \alpha(V_G - V_T) - \beta V_{DS}), \tag{A.4}$$

Let us evaluate the integral in Eq.(A.1) for the two types of barrier, Eq.(A.2) and Eq.(A.3), respectively:

$$\int_{x_0}^{x_{turn}} \sqrt{2m_c(a - bx - E)}\, dx = \sqrt{2m_c(a - E)} \int_0^{\frac{a-E}{b}} \sqrt{1 - \frac{b}{a-E}x}\, dx =$$

$$\sqrt{2m_c(a - E)} \int_{\frac{b}{a-E}x_0}^{1} \frac{a-E}{b} \sqrt{1 - y}\, dy = -\frac{2}{3}\sqrt{2m_c(a - E)} \frac{a-E}{b}(1-y)^{\frac{3}{2}}\Big|_{\frac{b}{a-E}x_0}^{1} = \frac{2}{3}\sqrt{2m_c} \frac{(a-E)^{\frac{3}{2}}}{b}\left(1 - \frac{b}{a-E}x_0\right)^{\frac{3}{2}} \quad (A.5)$$

For the case of the potential (A.2), $a = E_{min}$, $b = \frac{1}{L}(E_{min} + |qV_{DS}|)$, $x_0 = 0$. Since here $0 \leq E \leq E_{min}$, we can rewrite the last equality in Eq.(A.5) as follows:

$$P = \exp\left(-\frac{4}{3\hbar}\sqrt{2m_c}L \frac{(E_{min}-E)^{\frac{3}{2}}}{E_{min}+|qV_{DS}|}\right) \quad (A.6)$$

For the case of the potential (A.3), we obtain:

$$P = \exp\left[-\frac{2}{\hbar}\left(\sqrt{2m_c(E_b - E)}(L - \Lambda) + \frac{2}{3}\sqrt{2m_c}\frac{(a-E)^{\frac{3}{2}}}{b}\left(1 - \frac{b}{a-E}x_0\right)^{\frac{3}{2}}\right)\right]. \quad (A.7)$$

Here $a = E_b - \frac{E_b+|qV_{DS}|}{\Lambda}(\Lambda - L)$, and $b = \frac{E_b+|qV_{DS}|}{\Lambda}$, $x_0 = L - \Lambda$. So, we can write (A.7) as:

$$P = \exp\left[-\frac{2}{\hbar}\left(\sqrt{2m_c(E_b - E)}(L - \Lambda) + \frac{2}{3}\sqrt{\frac{2m_c}{\Lambda}} \frac{((E_b+|qV_{DS}|)L - \Lambda E - \Lambda|qV_{DS}|)^{\frac{3}{2}}}{E_b+|qV_{DS}|}\left(1 - \frac{(E_b+|qV_{DS}|)(L-\Lambda)}{(E_b+|qV_{DS}|)L - \Lambda E - \Lambda|qV_{DS}|}\right)^{\frac{3}{2}}\right)\right] \quad (A.8)$$

Here $E_b$ is given by Eq.(A.4), and we can regard that $E = 0$. We plot the dependence of the values given by Eqs.(A.7)-(A.8), on the voltages $V_G - V_T$ and $V_{DS}$, in Mathematica 12.2 software (https://www.wolfram.com/mathematica). Results are shown in **Fig. A1-2**, and in the main text.

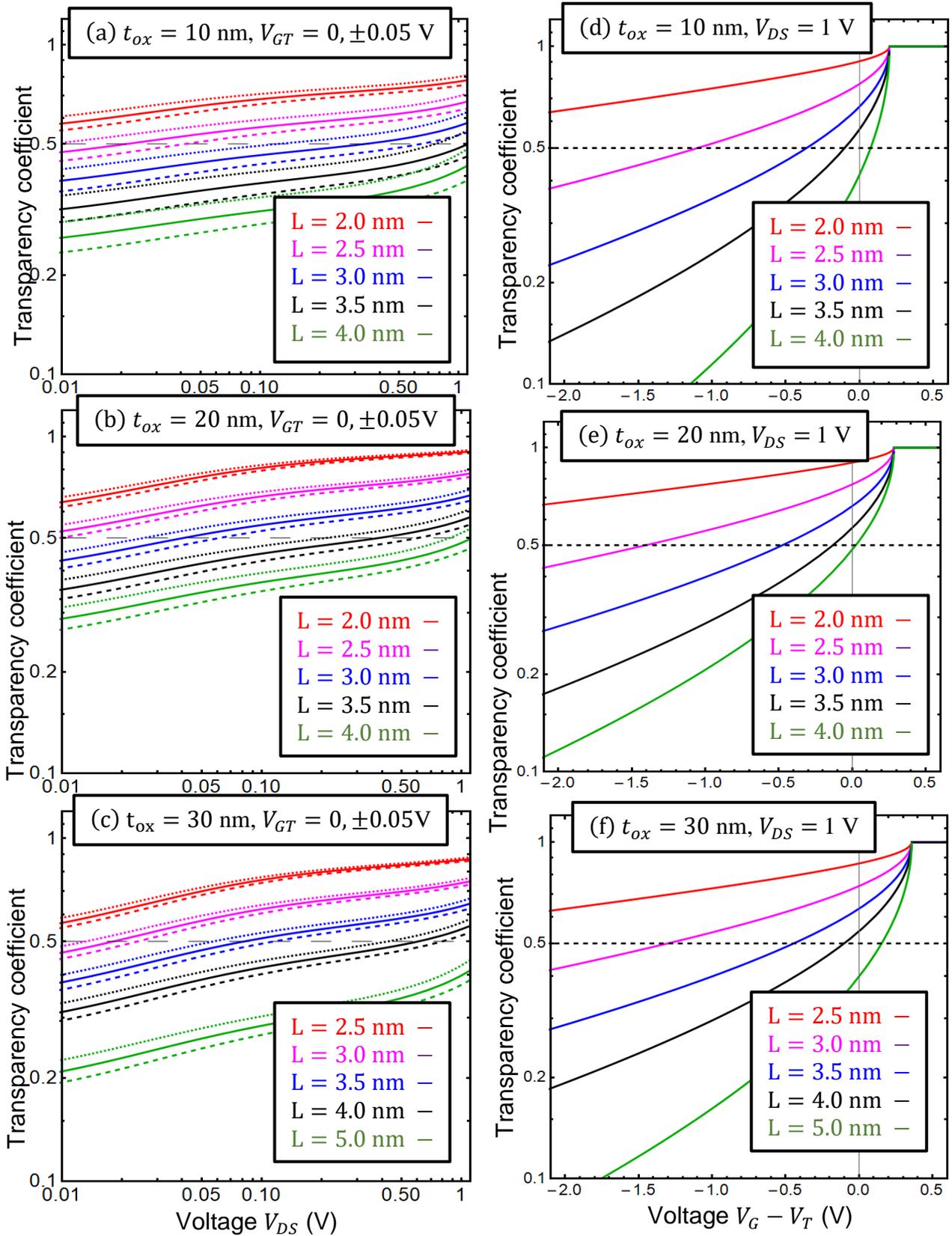

**Figure A1.** The dependence of the transparency coefficient on the drain-source voltage $V_{DS}$ for fixed $V_{GT} = -0.05\,V$ (dashed curves), 0 (solid curves) and 0.05 V (dotted curves). **(a-c)** and on the difference $V_G - V_T$ for the fixed $V_{DS} = 1$ V **(d-f)**. The different coloured curves correspond to the different barrier lengths $L$, namely $L = (1-5)$nm (listed in the legends) for $t_{ox} = 10$ nm, $\Lambda = 1.18$ nm and $\alpha = 0.069$ **(a,d)**; for $t_{ox} = 20$ nm, $\Lambda = 1.67$ nm and $\alpha = 0.051$ **(b,e)**; and for $t_{ox} = 30$ nm, $\Lambda = 2.05$ nm and $\alpha = 0.04$ **(c,f)**. Other parameters are listed in **Table AI**.

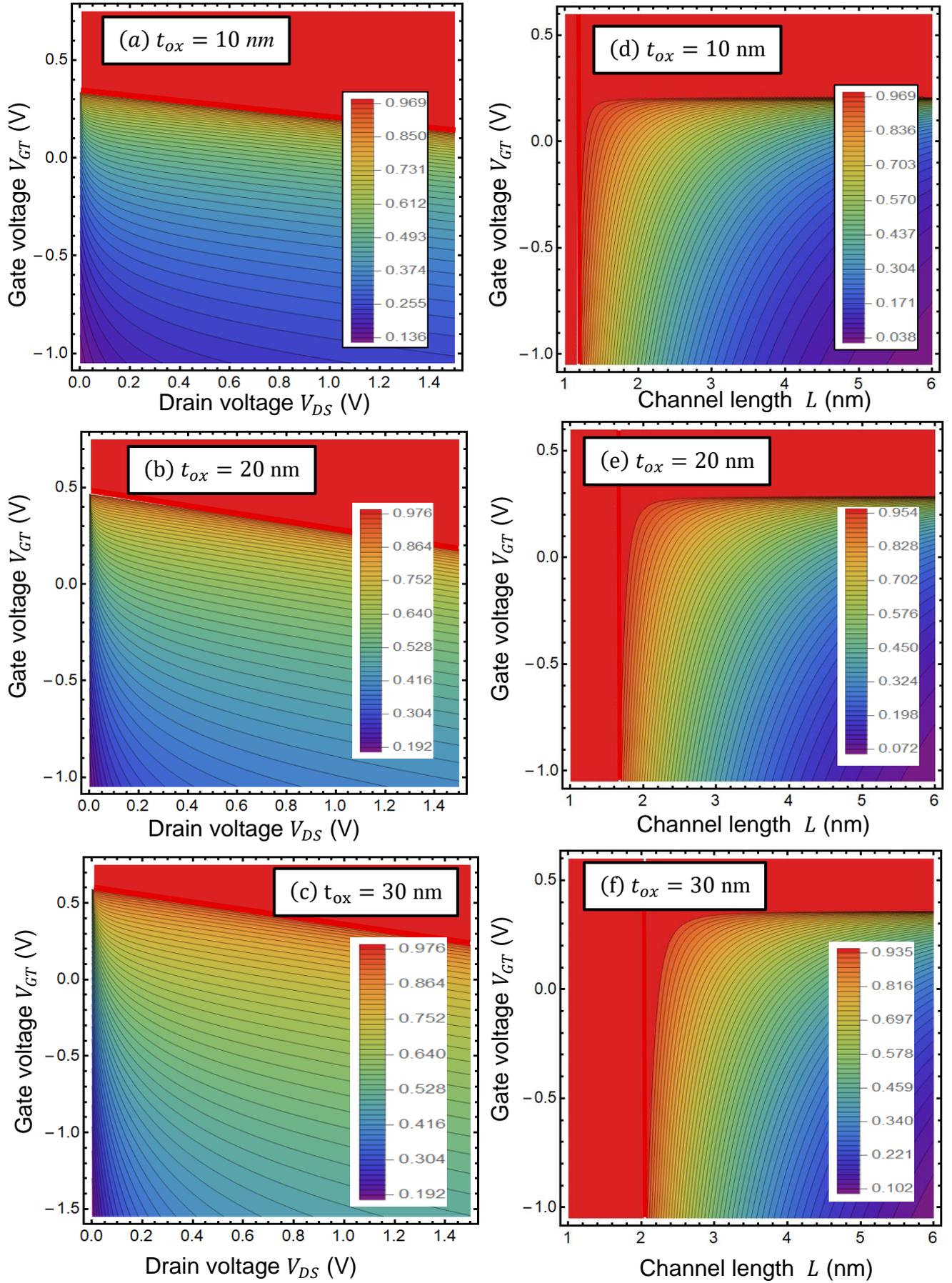

**Figure A2. (a-c)** The transparency coefficient vs. the drain-source and gate voltages, $V_{DS}$ and $V_{GT}$, for $L = 3$ nm. **(d-f)** The transparency coefficient vs. the gate voltage $V_{GT}$ and channel length $L$ for $V_{DS} = 1$ V. The parameters $t_{ox} =$

10 nm, Λ = 1.18 nm and α = 0.069 (**a,d**); for $t_{ox}$ = 20 nm, Λ = 1.67 nm and α = 0.051 (**b,e**); and for $t_{ox}$ = 30 nm, Λ = 2.05 nm and α = 0.04 (**c,f**). Other parameters are listed in **Table AI.**

**Table A1.** The parameters used in the calculations.

| $E$ | $V_{FB}$ | $E_b^0$ | $\psi_s$ | $\alpha$ | $\beta$ | T | $\varepsilon_s$ | $\varepsilon_{ox}$ | $t_s$ | $t_{ox}$ |
|---|---|---|---|---|---|---|---|---|---|---|
| 0 | 250 mV | 25 meV | 30 mV | 0.07-0.04 | 0.01 | 293 K | 5 | 25 | 0.7 nm | (10 – 30) nm |

$q$ and $m_0$ are the free electron charge and mass, $m_c = 0.55 m_0$, *HfO₂ lattice constant ~0.5 nm

# REFERENCES

[1] K. S. Novoselov, A.K. Geim, S.V. Morozov, D. Jiang, Y. Zhang, S.V. Dubonos, I.V. Grigorieva, and A.A. Firsov Electric Field Effect in Atomically Thin Carbon Films, Science **306**, 666 – 669 (2004).

[2] Yu.O. Kruglyak, M.V. Strikha. The Landauer-Dutt-Lundstrom model in the state of graphene's transportation to transport sites is used. Ukrainian Physical Journal. Reviews. (2015) - v. 10, No. 1. - P.3-32

[3] Li, X., Wang, X., Zhang, L., Lee, S. & H. Dai. Chemically derived, ultrasmooth graphene nanoribbon semiconductors. Science **319**, 1229 (2008).

[4] L. Jiao, L. Zhang, X. Wang, G. Diankov, & H. Dai, Narrow graphene nanoribbons from carbon nanotubes. Nature **458**, 877–880 (2009).

[5] F. A. Rasmussen and K. S. Thygesen. Computational 2D Materials Database: Electronic Structure of Transition-Metal Dichalcogenides and Oxides. J. Phys. Chem. C, **119**, 13169 (2015)

[6] B. Radisavljevic, A. Radenovic, J. Brivio, V. Giacometti and A. Kis. Single-layer MoS₂ transistors. Nature Nanotechnology, **6**, 147 (2011).

[7] A. Nourbakhsh, A. Zubair, R. N. Sajjad, A. Tavakkoli K. G., W. Chen, S. Fang, Xi Ling, J. Kong, M. S. Dresselhaus, E. Kaxiras, K. K. Berggren, D. Antoniadis, and T. Palacios. MoS₂ Field-Effect Transistor with Sub-10 nm Channel Length. Nano Lett. **16**, 12, 7798 (2016)

[8] Y. Chen, X. Wang, P. Wang, H. Huang, G. Wu, B. Tian, Z. Hong, Y. Wang, S. Sun, H. Shen, J. Wang, W. Hu, J. Sun, X. Meng, and J. Chu. Optoelectronic Properties of Few-Layer MoS₂ FET Gated by Ferroelectric Relaxor Polymer. ACS Appl. Mater. Interfaces **8**, 32083 (2016).

[9] Pin-Chun Shen, C. Lin, H. Wang, K. Hoo Teo, and J. Kong. Ferroelectric memory field-effect transistors using CVD monolayer MoS₂ as resistive switching channel. Appl. Phys. Lett. **116**, 033501 (2020).

[10] W. Cao, W. Liu, J. Kang, and K. Banerjee. An ultra-short channel monolayer MoS₂ FET defined by the curvature of a thin nanowire. IEEE Electron Device Letters **37**, 1497 (2016).

[11] L. Xie, M. Liao, S. Wang, H. Yu, L. Du, J. Tang, J. Zhao et al. "Graphene-contacted ultrashort channel monolayer MoS₂ transistors. Advanced Materials **29** (37), 1702522 (2017).

[12] Yu.O. Kruglyak, M.V. Strikha. Physics of nanotransistors: the theory of MOSFET in the traditional cycle, the basis of the model of the virtual loop and the approach to the application. Sen'sorna elektronika i microsystem technologies. - 2019. - v. 16, No. 1. - pp. 7-40.

[13] Yu.O. Kruglyak, M.V. Strikha. Physics of nanotransistors: the background of the model of passing and the model of the virtual loop - the MVS-passing model. Sensor electronics and microsystem technologies. - 2020. - v. 17, No. 4. - P.4 -22.

[14] M.V. Strikha, A.I. Kurchak. About the fundamental boundaries for the improvement of the MOSFET conduction channel with the real value of the bar potential. Ukrainian Journal of Physics (accepted).

[15] M. V. Strikha, A. I. Kurchak, and A. N. Morozovska. Fundamental constraints for the length of the MOSFET conduction channel based on the realistic form of the potential barrier (http://arxiv.org/abs/2012.11203)

[16] M. Lundstrom, Fundamentals of Nanotransistors (Singapore: World Scientific: 2018); www.nanohub.org/courses/NT

[17] K. K. H. Smithe, Chris D. English, Saurabh V. Suryavanshi, and Eric Pop. High-Field Transport and Velocity Saturation in Synthetic Monolayer $MoS_2$. Nano Lett. **18**, 4516−4522 (2018)

[18] H. Falco, T. Olsen, and K. S. Thygesen. How dielectric screening in two-dimensional crystals affects the convergence of excited-state calculations: Monolayer $MoS_2$. Physical Review B **88**, 245309 (2013).

[19] F. A. Rasmussen and K. S. Thygesen. Computational 2D Materials Database: Electronic Structure of Transition-Metal Dichalcogenides and Oxides. J. Phys. Chem. C, **119**, 13169 (2015)

[20] Yu.O. Kruglyak, M.V. Strikha. Physics of nanotransistors: devices, metrics and keruvannya. Sensor electronics and microsystem technologies. - 2018. - v. 15, No. 4. - S.18-40

[21] A. Majumdar, D. A. Antoniadis, IEEE Trans. Electron Dev., **61**: 351 – 358 (2014).

[22] I. Wakabayashi, S.Yamagami, N. Ikezawa, Ogura, Atsushi; Narihiro, Mitsuru; Arai, K.; Ochiai, Y.; Takeuchi, K.; Yamamoto, T. Mogami, T. (December 2003). IEEE International Electron Devices Meeting (2003): 20.7.1–20.7.3. doi:10.1109/IEDM.2003.1269446

[23] R. Landauer, IBM J. Res. Dev., **5**, 183 – 191 (1961).

[24] A. D. Franklin, M. Luisier, Shu-Jen Han, G. Tulevski, C. M. Breslin, L. Gignac, M. S. Lundstrom, W. Haensch, Nano Lett., **12**, 758 – 762 (2012)